\documentclass{elsart3p} 

\usepackage{graphicx}
\journal{Physics Letters B} 
\begin{document}
\begin{frontmatter} 
\title{Low-energy dipole strength and the critical case of $^{48}$Ca\thanksref{dfg} 
} 
\thanks[dfg]{Work supported by: the  
Deutsche Forschungsgemeinschaft 
through SFB 634; 
the Helmholtz International Center for FAIR 
within the framework of the LOEWE program launched by the state of Hesse;  
the BMBF Verbundforschung, contract 06DA9040I;   
the National Science Foundation, Grant No. PHY-0758125; 
and the UNEDF SciDac Collaboration under the U.S. Department of Energy Grant No. DE-FC02-09ER41585. 
\\ 
$^{\dagger}$ Present address: Dept. of Physics, Ohio State University 
} 

\author[ikp]{P.~Papakonstantinou\corauthref{cor}}  
\corauth[cor]{Corresponding author. 
Present address: IPN Orsay. Email: papakonst@ipno.in2p3.fr}
, 
\author[msu]{H.~Hergert$^{\dagger}$}, 
\author[ikp]{V.Yu.~Ponomarev}, 
\author[ikp]{R.~Roth}

\address[ikp]{Institut f\"ur Kernphysik, 
Technische Universit\"at Darmstadt,  
D-64289 
Darmstadt, Germany  
} 
\address[msu]{National Superconducting Cyclotron Laboratory, Michigan State University, East Lansing, Michigan 48824, USA} 


\begin{abstract} 
Recent theoretical work has not led to a consensus regarding the 
nature of the low-energy E1 strength in the $^{40,44,48}$Ca isotopes, for which high-resolution $(\gamma,\gamma')$ data exist. 
Here we revisit this problem using the first-order quasi-particle random-phase approximation (QRPA) and different interactions. 
First we examine all even Ca isotopes with $N=14-40$. 
All isotopes are predicted to 
undergo dipole transitions at low energy, of large and comparable isoscalar strength but of varying $E1$ strength.  
Provided a moderate and uniform energetic shift is introduced to the results,  
QRPA with the Gogny D1S interaction is able to account for the $(\gamma , \gamma')$ data, 
because, up to $N=28$, 
it yields a rather pure isoscalar oscillation. A neutron-skin oscillation is anticipated for $N\geq 30 $. 
This contradicts existing predictions that $^{44,48}$Ca develop a neutron-skin mode. 
Which theoretical result is correct cannot be resolved conclusively using the available data. 
We propose that alpha-scattering, possibly followed by an electroexcitation experiment, could resolve the situation and thereby help to improve the different models aspiring to describe reliably the low-energy dipole strength of nuclei. 
\end{abstract} 

\begin{keyword}
low-energy dipole stength; 
Ca isotopes; 
quasiparticle RPA and second RPA; 
electron scattering; 
alpha scattering; 
phenomenological and realistic Hamiltonians; 
\end{keyword} 

\end{frontmatter} 

\section{Introduction} 

Recent theoretical work \cite{Har2004,Ter2007,GGC201X} has led to a partial understanding, at best, of the 
nature of the low-energy E1 strength observed in Ca isotopes \cite{Har2004}. 
Effects beyond first-order random-phase approximation (RPA), or quasi-particle RPA (QRPA), seem necessary to reproduce certain details of the strength distributions, but contradictory reports exist as to whether a collective transition is present at all in $^{48}$Ca~\cite{Ter2007,GGC201X}. 
Here we show that QRPA  with a conventional force, Gogny D1S, also reproduces the observed systematics 
and provides a new  physical interpretation for it. 
We then propose a strategy to resolve which theoretical scenario is correct.  

In the following we avoid the loaded term ``pygmy dipole" strength, or resonance. 
We instead refer to electric-dipole transitions below the giant dipole resonance (GDR) simply as low-energy dipole states, or LEDs, regardless of their possible nature, 
and to their strength, especially E1, as low-energy dipole strength, or LED strength. The acronym IS-LED stands for isoscalar low-energy dipole states or strength. 
To refer specifically to an oscillation of a neutron skin against a rather inert core, we use the term ``neutron-skin mode", or NSM. 

Let us first recapitulate the current situation. 
High-resolution $(\gamma,\gamma')$ experiments have measured the E1 strength at excitation energies below 10 MeV in the three isotopes, $^{40,44,48}$Ca~\cite{Har2004}: rather little in $^{40}$Ca, and roughly 10 times more in $^{44}$Ca, $^{48}$Ca. 
That there is comparable and even somewhat less strength in $^{48}$Ca than in $^{44}$Ca is viewed as a puzzle, because it contradicts the simplistic expectation, founded on early calculations~\cite{CZA1994}, that LED strength should increase with neutron/proton asymmetry. 
The above reasoning disregards the fact that the strongest LED state in the $N=Z$ nucleus $^{40}$Ca carries too much E1 strength compared to other nuclei of similar mass~\cite{EnV1974}.  
Even so, RPA tends to predict too much E1 strength for this state~\cite{PPR2011}. 
For a useful comparison between theory and experiment 
one must therefore consider $^{40}$Ca to be part of the systematics, not just a core with negligible LED strength.

We point out that 
Ca isotopes cannot be considered too light to develop collective LEDs. 
Precisely such a mode is observed in $^{40}$Ca (and even in $^{16}$O) -- see Ref.~\cite{PPR2011}, where a compelling case was made for its coherent nature. 
If $^{48}$Ca does not develop a collective LED mode at all, its mass is not the underlying reason.  
Whether or not it is too light to develop a NSM~\cite{GGC201X}, is a different question. 
We stress that a NSM is not the only collective dipole vibration that could develop below the GDR and therefore not the only conceivable collective mechanism for generating LED strength. 
This becomes obvious in the case of $^{40}$Ca~\cite{PPR2011} and was stressed also in Ref.~\cite{Urb201X}, devoted to the toroidal dipole mode, and implied in Ref.~\cite{Bas2008}.

In general, the first-order RPA and QRPA 
cannot describe precisely the LED strength observed in Ca isotopes.  
In self-consistent (Q)RPA calculations, the lowest dipole states tend to lie too high in excitation energy,  
or the systematics tends to be wrong: 
E1 strength is predicted to rise almost linearly with neutron number $N$~\cite{CZA1994}, apparently due to the development of a neutron skin.  
Moreover, as already mentioned, the LED strength of $^{40}$Ca is often overestimated.  
The so-called Extended Theory of Finite Fermi Systems (ETFFS), which goes beyond RPA, is actually the only model which has reproduced the properties of LED strength observed in $^{40}$Ca, $^{44}$Ca and $^{48}$Ca quite accurately~\cite{Ter2007}.

Configurations beyond first-order (Q)RPA can affect the LED strength by   
1) shifting strength to lower energies, where LED strength is experimentally observed, as first suggested in \cite{GoK2002}; 
2) introducing fragmentation and possible quenching of strength; 
3) introducing additional states of two-phonon character, not described by RPA; 
4) redistributing strength so that much LED strength remains above the measurement endpoint of 10 MeV. 
All the above mechanisms have been touched upon in Refs.~\cite{Har2004,Ter2007,GGC201X}. 
A strong statement in Ref.~\cite{Ter2007} 
was indeed that in the case of $^{48}$Ca phonon coupling shifts a strong LED state resembling a NSM to energies higher than 10~MeV and thus the experiment missed it.  
This was proposed as an explanation of the observed systematics. 
Notably, an early Coulomb-excitation experiment found no evidence of a NSM mode below 12~MeV in $^{48}$Ca~\cite{Ott1999}. 

An important concern remains then that RPA-based models predict collective LED vibrations for all Ca isotopes, of NSM type or other, which have not been identified except for $^{40}$Ca. 
Obviously, RPA, like any model, cannot be expected to account for all possible collective nuclear states. 
The converse is not true, however: 
A collective RPA vibration that remains incomprehensibly elusive is not a minor problem. 
 
At this point we note that, if mechanism 1) above is a dominant one, then a reasonable description of LED strength should be possible within (Q)RPA, once a uniform energetic shift is introduced. 
We will make such an attempt in this work.  

We may now give an outline of the present work. 
First we present results with different effective interactions, Gogny and UCOM-based, for the IS and E1 strength distributions of even Ca isotopes with $N=14-40$ 
and discuss their general features. 
We then make an attempt to describe the experimental systematics within QRPA, by introducing an energetic shift as already mentioned.  
We find it successful when the Gogny D1S interaction is used. 
The reason is that, 
for $N$ up to 28, a predominantly isoscalar state is predicted, as opposed to a NSM.  
Such a scenario contradicts the ETFFS predictions~\cite{Ter2007}. 
We suggest that an electroexcitation experiment on $^{48}$Ca, preceded by alpha scattering, could resolve the situation.

\section{Theory} 
\label{S:theory} 

We employ the 
self-consistent Hartree-Fock-Bogolyubov--Quasiparticle-RPA (HFB-QRPA), 
which reduces to the self-consistent Hartree-Fock--RPA for closed-shell nuclei.  
The ground-state problem is solved within a single-particle basis spanning $15$ harmonic-oscillator shells. 
The same effective interaction is used to construct the QRPA equations, solved within the HFB basis. For details see \cite{HPR2011}. 
We consider a two-body Hamiltonian of the form 
\begin{equation} 
H = T + V_{\mathrm{NN}} + V_{\mathrm{Coul}} + V_{\rho} 
, 
\end{equation} 
where $T$ is the intrinsic kinetic energy, $V_{\mathrm{NN}}$ a two-body nuclear interaction, $V_{\mathrm{Coul}}$ the Coulomb interaction 
between protons and 
\begin{equation} 
V_{\rho} = t_3(1+x_3)\delta (\vec{r})\rho^{\alpha}(\vec{R}) 
\end{equation} 
is a density-dependent contact interaction ($\vec{r}$ the relative and $\vec{R}$ the center-of-mass position vector of the interacting nucleon pair). 
We employ  
the phenomenological Gogny D1S~\cite{BGG1991} parameterization 
and a unitarily-transformed AV18 realistic potential, supplemented with a phenomenological three-body contact term, 
which we label UCOM(SRG)$_{\mathrm{S,\delta 3N}}$. 
See Refs.~\cite{PPR2011,GRH2010} for information on the latter. 
For comparison we will also use 
Second-RPA (SRPA) for $^{40,48}$Ca, with the full coupling in the $2p2h$ space, and with 
the pure two-body UCOM-transformed AV18 potential, UCOM$_{\mathrm{var}}$ (UCOM-SRPA). 
This model is rather well suited for the description of the GDR region~\cite{PaR2009}. 
Here we introduce an energy cut-off of 140~MeV in the $2p2h$ space, 
so that the IS-LED state of $^{40}$Ca appears at approximately 7~MeV, i.e., close to its measured value.   

The ISD response is determined by the transition matrix elements of the operator 
\begin{equation} 
\hat{O}_{\mathrm{ISD}} = \sum_{i=1}^A e (r_i^3 - \frac{5}{3}\langle r^2\rangle r_i)Y_{1m}(\Omega_i )
\end{equation} 
and the electromagnetic response by  
\begin{equation} 
\hat{O}_{E1} = \frac{Ze}{A}\sum_{n=1}^N r_n Y_{1m}(\Omega_n ) - \frac{Ne}{A}\sum_{p=1}^{Z} r_p Y_{1m}(\Omega_p ) 
\end{equation} 
in an obvious notation, where the subscripts $p$ and $n$ refer to protons and neutrons, respectively. 
We calculate the excitation strength, $B(E1{\uparrow})$. 
The above operators include corrections to explicitly restore translational invariance. 
Within our self-consistent (Q)RPA calculations, we obtain practically the same values of strength 
if we use the uncorrected forms of these operators, 
except of course for the spurious state, which appears at practically zero energy. 
In SRPA spurious admixtures are unavoidable. 

Electroexcitation cross sections  
are calculated by using the proton transition density, $\delta\rho_p(r)$. 
For the longitudinal form factors in plane-wave Born approximation (PWBA) we use the convention 
\begin{equation} 
F_1(q^2) = \frac{\sqrt{12\pi}}{Z} \int_0^{\infty} \delta\rho_p (r) j_1(qr)r^2 \mathrm{d} r 
.
\end{equation} 
Within the distorted-wave Born approximation (DWBA), 
the cross section divided by the Mott cross section takes the place of the form factor squared. 

\section{Results} 
 \label{S:res} 
 
\begin{figure} 
\includegraphics[width=7cm]{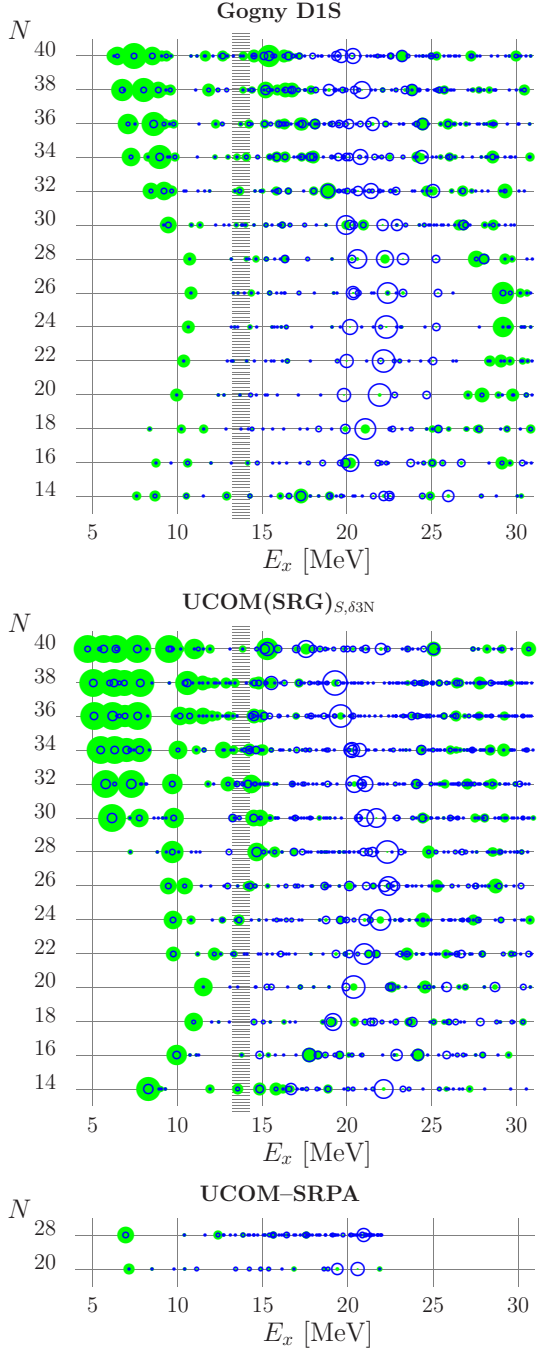}  
\caption{Isoscalar dipole (full circles) and E1 (open circles) strength in the Ca isotopes vs. neutron number $N$ and excitation energy $E_x$, 
for the Gogny D1S 
and the UCOM(SRG)$_{S,\delta 3\mathrm{N}}$ interactions, within QRPA, and for UCOM-SRPA (up to 22~MeV only). 
The areas of the circles are proportional to the strength. A different scale is used for the isoscalar and E1 strength. 
The shaded area corresponds to the energy region $13.25-14.25$~MeV (cf. fig.\ref{F:gamma}). 
\label{F:VsA}} 
\end{figure}

In fig.~\ref{F:VsA} we show the ISD and E1 strength distributions of the even Ca isotopes calculated 
using the Gogny D1S and the UCOM(SRG)$_{\mathrm{S,\delta 3N}}$ interactions. 
For $^{40,48}$Ca, UCOM-SRPA results are shown as well. 
The patterns and trends with respect to neutron number become clear in this mode of presentation, 
where the amount of strength at each value of excitation energy is represented by the area of a circle. 
A different scale is used for the IS (discs) and E1 (open circles) strength. 

The theoretical peaks of the GDR are visible as the larger open circles in fig.~\ref{F:VsA}. 
The GDR peaks of the $N=20-28$ isotopes are found experimentally at an excitation energy of, roughly, $18-21$~MeV.   
All three models reproduce this fact rather well. 
In all cases we find strong ISD states at the lower end of the spectrum. 
The same observation is made in a previous QRPA study using a different SRG-based interaction~\cite{HPR2011} and is in agreement with existing results using Skyrme functionals~\cite{CLN1997B,TeE2006}. 
These results are not surprising, because IS-LED states below threshold exhausting 4-14\% of the IS EWSR have been observed in various nuclei since years~\cite{HaV2001s1}. 
For $^{40}$Ca, the lowest-lying RPA state has been identified recently as the isospin-forbidden $E1$ transition~\cite{PPR2011}, experimentally observed at about 7~MeV~\cite{Gra1977,Poe1992}. 
Notably, the IS strength of the lowest strongly IS state of all isotopes and for both interactions is of the same order of magnitude, but the same does not hold for the E1 strength of those states. 

Let us examine the three sets of results separately, starting with QRPA and the Gogny D1S interaction.  
We first point out that the IS-LED strength is split in the neutron-deficient isotopes and even more fragmented for $N\geq 30$. 
For the lowest-lying peaks in the $N=18,20,22,24,26,28,30$ isotopes we find, respectively, 
590 (2 peaks below 10~MeV),      832, 860, 868, 855, 845, 1731 (two lowest peaks) {[}fm$^6${]} of IS strength and 
26.7 (two peaks), 4.5, 1.9, 1.6, 3.0, 6.0, 315 (two peaks) $[\times 10^{-3}e^2$fm$^2]$ E1 strength. 
The IS-LED strength doubles between  $^{48}$Ca and $^{50}$Ca and remains large for the heavier isotopes. 
The E1 strength increases fiftyfold 
from  $^{48}$Ca to $^{50}$Ca and remains in that order of magnitude in the heavier isotopes. 
At $N=30$ and beyond the energy of the lowest peak decreases. 
The E1 strength (but not the IS strength) reaches a minimum at $N=24$ and not at $N=20$. 
A similar counter-intuitive result appears in a recent study of shell effects in the isotopic behaviour of LED strength~\cite{INY201X}. 
The transition densities of these states reveal that they resemble a neutron-skin oscillation only for $N\geq 30$, not the lighter $N>Z$ isotopes.  
The NSM is of mixed isospin character and appears strong in both the IS and E1 channels. 
The stronger IS-LED states of the lighter $N>Z$ isotopes show a locally isoscalar character instead, resembling that of $^{40}$Ca, and therefore carry little E1 strength. 
In the case of $^{48}$Ca, for example, eight out of the 18 possible $1\hbar\omega$ $ph$ configurations (four of them proton configurations) contribute at least 1/18 each to the norm of the IS-LED state, making it quite a collective state~\cite{CDM2009}. 
Many $ph$ configurations contribute coherently to its IS strength. 
Dipole states in $^{48}$Ca with more-pronounced neutron transition densities are found at somewhat higher energies, but none can be viewed as an oscillation of a neutron skin against an inert core. 
Remarkably, the hole states in $^{48}$Ca are found very similar for protons and neutrons, a situation that is not realized in $^{40}$Ca.  
States dominated by proton transitions are found in the neutron-deficient isotopes. 
A variety of states with locally isoscalar transition densities, where both nucleon species contribute at the surface with non-negligible tails, are present regardless of $N$. 

The results for the  UCOM(SRG)$_{\mathrm{S,\delta 3N}}$ 
interaction are more in line with earlier predictions~\cite{CZA1994}: a minimum in E1 strength is reached for $^{40}$Ca, and a monotonic behaviour is observed on either side of $N=20$. 
At both $N=22$ and $N=30$ the position of the lowest peak shows a kink, pointing to structural changes of this state. 
Here a neutron-skin mode, characterized by a translational-type $\delta\rho_p(r)$ and an extended $\delta\rho_n(r)$ with a node, is predicted already for the lightest $N>Z$ isotopes. 
In the $N=20,24,28$ isotopes, however, these states carry too much E1 strength: 
they are about one order of magnitude stronger than the experimentaly observed transitions below 10~MeV~~\cite{Har2004}. 
The same holds for the results of Ref.~\cite{CZA1994}.  
Therefore the NSM scenario fails to explain the observed systematics, at the (Q)RPA level. 

Finally we look at the UCOM-SRPA results. 
As with QRPA and the UCOM(SRG)$_{\mathrm{S,\delta 3N}}$ interaction, we find  that the lowest strongly IS state alone carries too much E1 strength to be compatible with the data, namely 0.021 and 0.4 $e^2$fm$^2$ for $^{40}$Ca and $^{48}$Ca, respectively. 
In the case of $^{48}$Ca, this state is of NSM type and dominated by $(\nu f_{7/2})^{-1}$ transitions.  
Again the NSM scenario fails to describe the data. 

Within RPA with the 
UCOM(SRG)$_{S,\delta 3 \mathrm{N}}$ interaction 
as well as within the UCOM-SRPA model, 
we find that other IS states of $^{48}$Ca, at somewhat higher energies but still below the GDR, 
are of locally isoscalar character with a node in the transition densities of both nucleon species.   

From the present results we conclude that all isotopes develop strong IS-LED states, but the nature of these states changes as $N$ increases from resembling the 
oscillation of a proton skin, to a rather pure isoscalar oscillation, to a neutron-skin oscillation. 
The neutron number at which the latter transition takes place depends on the interaction. 
The energy and fragmentation patterns also change.  
An important observation is that E1 strength is found in the whole energy range from the lowest-lying state up to the GDR. 
Especially for the heavier isotopes, a clustering of strength in two or more energetic regions is apparent. 
Judging from the variety of the calculated transition densities, we expect that the mixing and fragmentation of different possible vibrational modes gives rise to this result: 
oscillations of a strongly decoupled neutron (or proton) 
skin against a core, or oscillations of a mixed layer of protons and neutrons against a core, 
these being of toroidal type or of partially compressional type -- see, for example, \cite{Bas2008,PPR2011,Urb201X,VWR2000,Rye2002,Kva201X} for studies of various possible dipole modes with different approaches.   
In general, the position of the torus or compression point may vary with energy and $N$. 
GDR admixtures and $1p1h$ transitions are also possible.  
A detailed study of all these states goes beyond the scope of this work. 
Nevertheless, we take the opportunity to 
point out again that E1 strength alone is not a good indicator of collectivity, in particular the lack thereof. 
A collective IS dipole state could in principle  carry less E1 strength than a $1p1h$ transition~\cite{PPR2011}. 
It may be telling that LED strength is often overestimated, even when it is interpreted as due to non-collective transitions~\cite{OHD1998}. 

\begin{figure} 
\centering\includegraphics[angle=-90,width=8cm]{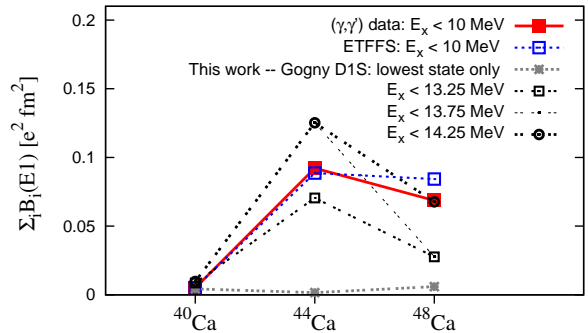}\\[2mm]  
\caption{Summed electric dipole strength for the three isotopes up to the indicated excitation energies: experimental data~\cite{Har2004}, ETFFS results~\cite{Ter2007} and present QRPA results with the Gogny D1S interaction. For the Gogny D1S model the E1 strength of only the lowest IS-LED state is also given. 
\label{F:gamma}} 
\end{figure}

We now return to our main topic, which is the LED strength of $^{40,44,48}$Ca, and ask whether the scenario predicted by the use of the Gogny D1S interaction is compatible with the $(\gamma ,\gamma')$ data. 
A revealing comparison is made in fig.~\ref{F:gamma}. 
Here we show the summed E1 strength observed below 10~MeV, the existing prediction within ETFFS, and our new sets of results: 
The E1 strength of the lowest dipole state in each isotope within (Q)RPA using the Gogny D1S interaction; the summed strength below 13.25~MeV, 13.75~MeV, and 14.25~MeV within the same model. 
The lowest states are too weak with respect to experiment. 
A similar observation was made in ref.~\cite{TeE2006}, where the Skyrme SkM* interaction was used, and in ref.~\cite{HPR2011}. 
The energy cut-off strongly influences the quality of the comparison. In fact, minor uncertainties in the data and the theoretical results can likewise affect the systematics.  
Therefore an attempt to reproduce the data very precisely in this region appears rather meaningless.  
Nevertheless, the measurements are rather well reproduced by QRPA for the higher cutoffs, indicated in fig.~\ref{F:VsA} by the shaded area. 
This generic observation remains true if we change certain ingredients of our calculation such as the harmonic-oscillator parameter. 
That it takes an energy shift by about $3-4$~MeV for the QRPA to describe the data, is in excellent agreement with the early calculations of Ref.~\cite{GoK2002}. 
Higher-order configurations are expected to shift the strength to lower energies and could introduce the fragmentation necessary to describe the E1 values more precisely~\cite{Ter2007,GGC201X,GoK2002,CoB2001,SBC2004}.  
We conclude that the scenario whereby 
a strong isoscalar mode with little E1 strength is present in $^{48}$Ca, and not a NSM, is very well 
compatible with the measurements. 
 
\section{Resolving the situation in $^{48}$Ca} 

We have seen that a strongly isoscalar state developing in $^{48}$Ca is as compatible with the existing data as a mixed-isospin NSM appearing above 10~MeV. 
We now ask how the correct scenario can be resolved.   
Measurements of the full dipole spectrum of $^{48}$Ca could be useful, because a NSM would be visible as a strong E1 transition close to particle threshold. 
Such information can be obtained from relativistic Coulomb excitation in proton scattering under extreme forward angles as recently demonstrated for the case of $^{208}$Pb \cite{Tam2011}. Data for $^{48}$Ca have been taken \cite{Tam2009} and are presently analyzed.
But if no such strong state is observed, the problem remains unresolved.  
We now note that both types of mode, namely pure IS or NSM, would carry much ISD strenght and therefore would feature prominently in $\alpha$-scattering. 
In particular, $(\alpha ,\alpha'\gamma)$ is very well suited to detect IS-LED strength below threshold~\cite{HaV2001s1,Zil2002,Sav2006,End2010}, 
but has not been used on Ca isotopes other than $^{40}$Ca~\cite{Poe1992}. 
Such an experiment on $^{48}$Ca would be most useful in resolving the present situation. 
Inelastic $\alpha-$scattering above threshold has been performed on $^{48}$Ca~\cite{Lui2011}. 
A resonance beyond 16~MeV was found, but no strong state was clearly observed in the energy region of the NSM mode as predicted by ETFFS.  The onset of another concentration of IS strength below 10~MeV could be inferred from the data, but not conclusively. 
 
\begin{figure} 
\centering\includegraphics[width=8cm]{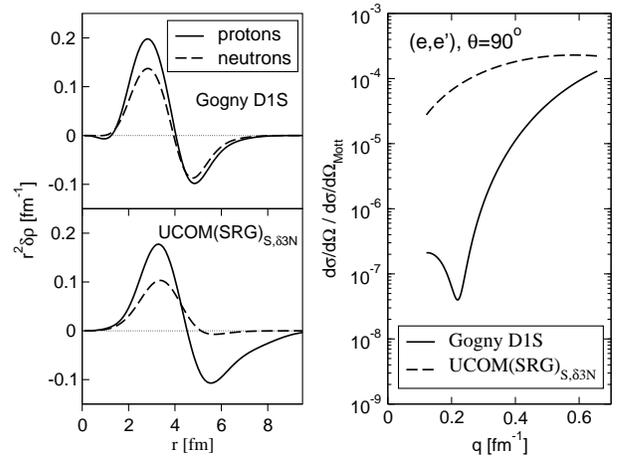} 
\caption{Transition densities and longitudinal electroexcitation cross section of the IS-LED state in $^{48}$Ca based on the Gogny D1S and the UCOM(SRG)$_{S,\delta3\mathrm{N}}$ interactions.  
\label{F:ff}} 
\end{figure}

If a strong isoscalar state is found below 10~MeV, it would likely not be of NSM type. 
An electroexcitation experiment could then reveal its nature.  
Inelastic electron scattering has been performed on $^{48}$Ca by different groups, but the analyses have mostly focused on $1^+$ states and other multipolarities.  
In fig.~\ref{F:ff} we show the transition densities of the IS-LED state as predicted 
within RPA by the Gogny D1S interaction and of the NSM mode as predicted, in this case, by UCOM(SRG)$_{S,\delta 3\mathrm{N}}$.  
We also show the corresponding longitudinal form factors within DWBA. 
The NSM has a large form factor already at low momentum transfer, while a completely different result is obtained for the other IS-LED state. 
A minimum at low momenta would make this state analogous to the IS-LED state of $^{40}$Ca~\cite{PPR2011}. 
Significant differences between the transverse form factors of the two different modes are also expected. A more dedicated study shall be the subject of upcoming work.  

\section{Conclusion} 

Our QRPA results suggest that all even Ca isotopes with $N=14-40$ develop strong IS-LED states. 
IS and E1 strength is present in the whole energy region up to the GDR. 
The nature of the states varies with the neutron number and  the energy. 
Results with the Gogny D1S interaction suggest that the lowest dipole transition becomes a neutron-skin mode only for $N\geq 30$. 
This result is compatible with the systematics of E1 strength observed in $^{40,44,48}$Ca, but contradicts an earlier interpretation of the data~\cite{Ter2007}. 
We propose that $\alpha -$scattering followed by an electroexcitation experiment on $^{48}$Ca could resolve the situation conclusively. 
We take the opportunity to stress that, in order to understand the nature of the whole low-energy dipole spectrum, one must examine not just the E1 strength, 
but at the same time the response to isoscalar operators, in particular toroidal and compressional ones~\cite{Kva201X}.  
These remarks are relevant for all isotopic chains and for problems including the nature of pygmy dipole strength in very neutron-rich nuclei, the so-called isospin splitting of pygmy dipole strength 
and the origin of E1 and isoscalar strength throughout the region below the GDR. 

{\bf Acknowledgments} 
We wish to thank Profs. P. von Neumann-Cosel for useful suggestions and A.Richter for related information.

\end{document}